\documentclass[conference]{IEEEtran}

\usepackage{graphicx}
\usepackage{amsfonts}
\usepackage{amssymb}
\usepackage{amsmath}
\usepackage{algorithm}
\usepackage{algorithmic}
\usepackage{amsthm}
\usepackage{hyperref}
\usepackage{acronym}
\usepackage{framed}
\usepackage{centernot}
\usepackage{color}
\usepackage{xcolor}
\usepackage{caption}
\usepackage{listings}
\usepackage{tcolorbox}
\usepackage{lipsum}
\usepackage{drawstack}
\usepackage{tikz}
\usepackage{float}
\usetikzlibrary{tikzmark}
\usetikzlibrary{positioning}

\tikzset{set/.style={draw,circle,inner sep=0pt,align=center}}

\definecolor{mygreen}{rgb}{0,0.6,0}
\definecolor{mygray}{rgb}{0.5,0.5,0.5}
\definecolor{mymauve}{rgb}{0.58,0,0.82}

\ifCLASSINFOpdf
\else
\fi
\begin{document}
%
\title{Mission Impossible: Securing Master Keys}

\author{\IEEEauthorblockN{Hannes Salin}
\IEEEauthorblockA{Falun, SWEDEN\\
hannessalin@gmail.com}
\and
\IEEEauthorblockN{Dennis Fokin}
\IEEEauthorblockA{Stockholm, SWEDEN\\
fokin.dennis@gmail.com}}


%


\newtheorem{mydef}{Definition}
\newtheorem{mythm}{Theorem}
\newtheorem{myprop}{Proposition}
\newtheorem{mylem}{Lemma}
\newtheorem{myex}{Example}
\newtheorem{myalgorithm}{Algorithm}
\newtheorem{mysolution}{Solution}
\newtheorem{myprinciple}{Principle}

\maketitle

\begin{abstract}
Securing a secret master key is a non-trivial task, we even argue it is impossible to fully secure it, hence we must make it as difficult as possible for any powerful adversary to steal or use the key. We introduce the reader to interesting cryptography which is starting to get more attention in terms of addressing the above problem, and we briefly overview some commercial and open-source products that can be used. Finally, we propose a set of solutions on how to secure master keys, more as guidelines rather than exact technical specifications, with aim to inspire and raise awareness of how to increase the security as much as possible.
\end{abstract}


%
\IEEEpeerreviewmaketitle


\section{Introduction}
Given a master key that can unlock all your secrets and derive any encryption key you may have in your system, how would you securely store and hide that key? You can encrypt it, but then all security relies on another secret key or password used when encrypting the master key. You may put it in a Hardware Security Module (HSM) or a vault but again it would only be protected by another key or password. It seems that no matter which approach is used for securing a secret key, you must use yet another key, hence the problem is only moved one layer ahead - we can suspect this problem unsolvable!
\\
\\
Assuming that no fully secure way to protect a secret key exists, we can only do our best to make it as difficult as possible for an adversary to break our system and steal the key (or at least get access and make use of the key). 
Securing secret keys is a pretty normal concept in modern software development, especially in web applications, API:s and microservices that utilizes TLS or integrates to other systems using secure channels. Due to the nature of secure communication, we need all of these private and secret keys to establish and perform symmetric encryptions and digital signatures. 
It is not too uncommon to find that the secret key is embedded in the service itself, sometimes in plain text and sometimes in a keystore. We would argue it is pretty common to use property- or configuration files to point to a keystore or the actual key, which the service then reads to memory when requested to perform encryption.
If an adversary manages to compromise the underlying server or memory, it would be theoretically possible to steal that key, or if a vault is used, then the adversary would be able to intercept the password or key used for unlocking the vault.
\\
\\
Note that we are assuming systems where the usage of secret keys are automated and not manually handled by a human; in those cases there might be a reason for a person to actually input a device and/or typing a secret password to unlock a master key or other highly sensitive data. 
\\
\\
Upcoming sections will briefly overview current techniques and crypto protocols for securing master secrets, both on what is adopted into commercial products and discovered in recent research. Afterwards, we are ready to conclude best solutions for storing master keys and propose a set of solutions based on secret sharing with multi-party computations. 

\section{Securing secret keys}

For larger organizations it is standard practice to use HSMs for storing secret keys and performing cryptographically important operations such as digital signatures or generating symmetric keys, especially if there is a demand for high level security. An HSM is basically a piece of hardware, specifically constructed for crypto operations or secure storage of data, which often is hardened and tamper-resistant. Usually the HSM is FIPS-140 certified as well, meaning fulfilling US government requirements for cryptography modules \cite{brown1994security}.

Popular commercial HSM vendors are Thales \cite{thales} and Utimaco \cite{utimaco} who offer a wide range of different models and functions. Depending on the setup of the HSM, or rather the cluster of HSMs for redundancy, applications request a signature or encryption of a message $m$ using the master key inside the HSM. Typically $m$ is sent from the application to the HSM along with some access such as password or secret key, and if authenticated correctly the HSM will perform the requested crypto operation. Moreover, the HSM can be accessed for operational or maintenance purposes, then a secure channel is needed and a specific administration account can be used for login. For Thales SafeNet HSM this is typically done by using an admin account with a secret PIN code.
\\
\\
If not using an HSM it is common to use a virtual safe, often referred to as a \emph{vault}.

\section{Secret Sharing}
The purpose of secret sharing is for a party $P_i$ to spread a secret across other parties in such a way that only $P_i$ can recreate the secret. As an example, assume that a party $P_1$ wants to share a secret integer $x$ in $\mathbb{Z}_p$, where $p$ is a prime. By finding three other integers such that
$$
x = r_1 + r_2 + r_3 
$$
it is possible for $P_1$ to secretly share $x$ with two other parties $P_2, P_3$. This is accomplished by sending $r_1$ and $r_3$ to $P_2$, $r_1$ and $r_2$ to $P_3$, and $r_2$ and $r_3$ to $P_1$ itself. $P_1$ has successfully managed to secretly share the secret. Neither party alone can compute $x$, but it can be reconstructed by using shares from at least two parties \cite{cramerbook}. One of the more widely used secret sharing schemes was created by Adi Shamir and is called Shamir's secret sharing \cite{shamir1979share}, which we will explain in greater detail below. However, it is worth pointing out that other schemes exists, for example Blakley's secret sharing \cite{blakley1979safeguarding}. Secret sharing is a key tool used in secure multi-party computation.

\subsection{Shamir's secret sharing}
As stated by Shamir \cite{shamir1979share}, a $(k,n)$-threshold scheme is a scheme where a secret $D$ is divided into $n$ shares (or pieces): $D_1,...,D_n$ such that:
\\
\begin{enumerate}
\item{$D$ is easily computed by knowing $k$ or more pieces, and}
\item{$D$ cannot be computed by knowing $k-1$ or less pieces.}
\end{enumerate}
\ \\
The idea behind a threshold scheme is that they are helpful for managing cryptographic keys. Shamir states that in order to protect data we encrypt it. However, it is not possible to encrypt the keys themselves, since these are needed for encryption and decryption. Thus, Shamir proposes a robust key management scheme by using a $(k,n)$-threshold scheme where $n = 2k-1$. This way the key can be recovered even if $n/2 = k-1$ shares are destroyed \cite{shamir1979share}.

The scheme Shamir proposed is based on polynomial interpolation. Given $k$ points in the 2-dimensional plane $(x_1,y_1), ..., (x_k, y_k)$ there is only one polynomial $q(x)$ of degree $k-1$ such that $q(x_i) = y_i$ for all $i$, given that all $x_i$ are distinct \cite{shamir1979share}.

In order to divide the secret $D$ into $n$ shares, a random $k-1$ degree polynomial $q(x)$ is chosen, where $q(x) = a_0 + a_1x + a_2x^2 + ... + a_{k-1}x^{k-1}$ and $a_0 = D$. The share $D_i$ is then calculated by evaluating:

$$
D_1 = q(1), ..., D_n = q(n)
$$

Given any subset with $k$ shares, the coefficients of $q(x)$ can be found by interpolation. The secret is found by evaluating $q(0) = D$ \cite{shamir1979share}.

\subsection{Verifiable secret sharing}
A problem with secret sharing is that any of the involved parties might become malicious, for example the dealer or any of the participating parties. In the former case, the dealer would send incorrect shares to the other parties. In the latter case, a bad party might interfere with the protocol and send a faulty share to the dealer. This would break the protocol, since even if there are enough good players to recover the secret, no one can distinguish good shares from faulty ones. In order to solve this problem, verifiable secret sharing schemes can be used instead, where each party can verify that their share is consistent. Paul Feldman \cite{feldman1987practical} proposed a modification of Shamir's protocol utilizing homomorphic encryption in order to make it verifiable.

\section{Secure Multi-Party Computation}
Assume two parties want to find out who is richer, without revealing their wealth to each other. How would you solve that? Andrew Yao \cite{yao1982protocols} answered this question in 1982, thus creating a subfield within cryptography known as \emph{secure multi-party computation} (SMPC). The basic idea is that a set of parties keeps a bit of a secret each, which all (or at least a threshold thereof) is needed when restoring the original secret and input it to some secure function which computes a desired outcome everyone are interested in. Therefore all parties will not reveal their own secrets but can take part of a common output, e.g. the answer to who is richest.

Nowadays with increased utilization of cloud services and distributed infrastructures, SMPC-based protocols could be something both relevant and focused on if theory can be bridged over to secure and stable implementations. Also, from Garner's Hype Cycle for security in 2019, multi-party computations are put under the category \emph{on the rise} \cite{gartner} which indicates that this technology might be something to start look into more seriously.
\\
\\
In a SMPC problem there are $n$ players, or parties, that participate: $P_1, P_2, ..., P_n$. Each player $P_i$ holds a secret input $x_i$ which they do not want to reveal to any of the other players. The players agree on a function $f$ that takes $n$ parameters as input and outputs an answer $y$, such that $y = f(x_1, x_2, ..., x_n)$. At the same time, two conditions must be satisfied:
\\
\begin{enumerate}
\item{\emph{Correctness}: the correct value of $y$ is computed.}
\item{\emph{Privacy}: $y$ is the only new information released.}
\end{enumerate}
\ \\
If both condition 1) and 2) are satisfied when computing function $f$, it is said that $f$ has been computed securely \cite{cramerbook}.
\\
\\
A trivial solution to this problem is that all parties agree on a trusted third party $\mathcal{T}$ which would gather all the inputs and compute the function $f$ \cite{cramerbook}. $\mathcal{T}$ announces the result to the parties and then discards all computation results. As pointed out by Cramer et al \cite{cramerbook} there are at least two problems with this approach. Firstly, it creates a single point of failure. If an adversary gets access to the third party, all private data could be stolen. Secondly, all parties must now trust $\mathcal{T}$. If they do not even trust each other, how are they supposed to trust $\mathcal{T}$?

\subsection{SMPC protocols and homomorphic encryption}
Some SMPC protocols are combined with \emph{homomorphic encryption} (HE),  i.e. encryption schemes where you can compute on ciphertexts without decryption. This is particularly interesting since then all parties $P_i$ can encrypt their inputs under some HE scheme, next they evaluate function $f$ using the ciphertexts and finally execute the distributed decryption to get the result. The problem is that in theory so called \emph{fully homomorphic encryption} (FHE) is needed but is in practice too slow for real implementations \cite{damgaard2012multiparty}. The field is still in its infancy but some progress on the practical development is initiated, e.g. IBM has released an open-source library for various forms of homomorphic encryption \cite{halevi2013design} and Inpher has its open-source library called \emph{TFHE: Fast Fully Homomorphic Encryption over the Torus} \cite{tfhe}.
\\
\\
Rivest et al. proposed in the late 70s \cite{fontaine2007survey} the idea of HE when trying to solve the problem of performing computations on encrypted data without decrypting it, thus it would be possible to run such operations on sensitive data on untrusted computers. More formally we could define an encryption scheme to be \emph{homomorphic} if 
$$
\forall m_1, m_2 \in \mathcal{M}, E(m_1 \odot_\mathcal{M} m_2) = E(m_1) \odot_\mathcal{C} E(m_2)
$$

where $\mathcal{M}$ is the plaintext space and $\mathcal{C}$ ciphertext space respectively. Several other properties are also considered to gain different type of HE but the range of different schemes also have varying levels of security analysis \cite{fontaine2007survey}. More importantly we must differentiate \emph{fully homomorphic encryption} (FHE) which is the strongest notion meaning it would be possible to do arbitrary computations on the cipher texts; with weaker notions we loose the arbitrarily part. In any case, many SMPC protocols with FHE and variants thereof are proposed \cite{hui2017secure}, \cite{huang}, \cite{das} and \cite{damgaard2012multiparty}, and seemingly hints of increased interest of such protocols is to be found due to the increasing number of use cases to apply: \cite{song2020iris}, \cite{wei2020pairing}, \cite{wood2019homomorphic}, \cite{8962262} and more.

\section{Implementations and technology}
This section overviews commercial and open-source projects utilizing secret sharing, secure multi-party computations and similar. These are not too widely adopted yet but some commercial products outside of cryptocurrencies claim to use these algorithms and protocols; we often see that cryptocurrencies also uses more and other type of cryptographical primitives such as zero-knowledge proofs \cite{sanchez2019zero}, accumulators \cite{6547123} and others. 
\subsection{Docker secrets}
Using containerization is common today and Docker \cite{docker} is one of the more well known technologies adopted. Having a cluster of Docker containers, possibly communicating among each other or with external services, the need for storing private keys and other sensitive data is of importance. Therefore Docker built \emph{Docker secrets} \cite{dockersecrets} which handles any secret or sensitive data within a Docker ecosystem, or swarm. A pre-setup of a public key infrastructure is generated for all containers in a swarm, with purpose of being able to have mutually authenticated TLS with the \emph{swarm manager}. The manager is a node within the swarm and makes use of a Raft log \cite{ongaro2016search} to encrypt and store any sensitive data or keys which the container nodes needs. According to \cite{dockersecretsblog}, \cite{dockersecrets2} the encryption uses Salsa20Poly1305 with 256-bit keys. Furthermore, as mentioned in \cite{dockersecretsblog}:
\\
\\ 
\emph{"Secrets get saved to the internal Raft store using NACL’s Salsa20Poly1305 with a 256-bit key. When a swarm manager starts up, the encrypted Raft logs containing the secret is decrypted using a data encryption key that is unique per-node.
\\
Nodes cannot request access to the secrets themselves – they must be provided access by the swarm manager. When you grant access to a secret, it is sent over the already established TLS connection."}
\\
\\
To conclude we see that Docker solves the problem of accessing secrets by encrypting secret keys in a Raft log, which by nature builds on consensus algorithms meaning having several nodes work coherently (distributed) in a fault-tolerance context. Moreover, Salsa20Poly1305 cipher with 256-bit keys are used for encryption. Obviously the security boils down to the TLS part which each Docker node uses to access the secret from the swarm manager.

\subsection{Vault12}
Cryptocurrencies have become vastly popular since the release of Bitcoin and in order to protect such digital currencies, investors store them in hardware or software wallets. The founders of Vault12 \cite{vault12} argue that using these wallets for storing cryptocurrency assets has a major drawback in the case of availability. If the wallet is lost, so is some of the most valuable assets a person owns. Instead, they propose Vault12 - a new decentralized cryptographic security platform.

Vault12 utilizes Shamir's secret sharing to split critical data, such as keys, into $m$ number of shares, where $n$ shares are required to reconstruct the data. When creating the vault the owner recruits people to act as parties in the secret sharing scheme. Each party gets a share, including the owner. Their phones will also serve as distributed storage nodes. In order to unlock the vault the owner has to use social verification to request the shares from the other parties. Availability is strengthened by making sure that private keys (setup on the master device) are also divided into shares, however they are only sent to the other parties. In case the owner loses his master device, the owner can contact the parties and thus request restoration shares, used to restore the master device setup on a completely new phone.

\subsection{HashiCorp Vault}
HashiCorp offers a software solution vault \cite{vault} called Vault for storing and managing access to secret keys, tokens, passwords and certificates which in turn gives access to sensitive data. Vault makes it possible to manage these secrets centrally and with a cloud-based infrastructure in mind where nodes and systems are spread across platforms and networks. Furthermore, Vault can issue short-term certificates and act as a certificate authority. 

As with any type of vault solution there is also an associated access management layer; for Vault many integrations to cloud- and SAML-based solutions exists. This means that the product still relies on the authentication of the chosen method, e.g. AWS or Azure authentication, to secure the keys inside Vault. The protection against extraction is most likely high but with a broken authentication scheme an attacker would still be able to access the operations to be performed by the secret key. 

\subsection{Software development}
For system developers curious in exploring secret sharing and SMPC, several relevant programming libraries can be found. This list is by no means extensive but highlights somewhat active and usable resources available:
\\
\begin{itemize}
\item{Secret sharing}
  \begin{itemize}
  \item{\emph{Shamir secret sharing library}, written in Java, this is an implementation of the algorithm with a $(k,n)$-threshold that splits a secret $s$ into $n$ shares, where $k$ shares are needed in order to recover $s$. Lagrange interpolation is used to recover a secret. \cite{shamirjava}.}
  \end{itemize}
\item{Secure multi-party computations}
  \begin{itemize}
  \item{\emph{MPyC: Secure Multiparty Computation in Python} \cite{mpyc}, a python library originating from ideas developed from a PhD project at Aarhus university, Denmark and later as a postdoc project at Eindhoven University of Technology, Netherlands.}
  \item{\emph{EMP: Efficient Multi-Party computation Toolkit} \cite{emp-toolkit}, c++ based and includes both primitives and full protocols.}
  \end{itemize}
\item{Bilinear maps / pairings}
  \begin{itemize}
  \item{\emph{JPBC: Java Pairing-Based Cryptography library}, a library in Java used to perform pairing-based operations \cite{ISCC:DecIov11}.}
  \end{itemize}
\item{Homomorphic encryption}
  \begin{itemize}
  \item{\emph{HELib}, a C++ library using Brakerski-Gentry-Vaikuntanathan (BGV) scheme with a set of optimizations \cite{halevi2013design}.}
  \item{\emph{TFHE: Fast Fully Homomorphic Encryption over the Torus} \cite{tfhe}}, also a C/C++ library where a set of interesting implementations associated to Machine Learning and encryption can be found on the GitHub page for the project.
  \end{itemize}
\end{itemize}
\;
We cannot guarantee that any of the above mentioned resources are completely verified or reviewed on a satisfiable level but instead should be considered for inspiration.

\section{Solution proposal}
We structure our proposed solutions into two parts where non-critical and critical system are considered. The former is loosely defined as a system which does not require compliance to security standards or have any risk for high-loss implications if attacked. Normally such non-critical system should also be relatively low in cost when implementing thus the compromise between adequate security and user functionality is not highly relevant (for functionality's favor). We assume that low cost is also associated with no means of investing in a HSM.

Regardless of method for accessing a secret key or using a private key for secure communication, it must inevitably be read into memory at some point. Therefore, from a system development perspective it can be very useful to programatically make sure that any cryptographic material is cleaned out from memory after usage. How to do that differs depending on programming language and underlying platforms, but it should be taken into consideration that sensitive data might also spill over to swap and crash dump files. If possible, make sure that such storage is cleaned out as well. 

\subsection{Solutions for non-critical systems}
For non-critical systems it is sufficient of using keystores embedded in the application, or if possible using a Vault-based solution e.g. HashiCorp's Vault open-source version which is free. Most importantly, one should at least avoid storing any keys or passwords in plain text. It is not too uncommon that the application, for simplicity, reads a password from a configuration in order to access a keystore. An alternative is to read that password from a database instead, thus storing database credentials in the configuration. This is of course not a good thing either since the attacker get access to the database. On the other hand, it is at least possible to setup more mitigations on the database layer, e.g. restricted tables/schemes and access filtering on IP/MAC address and so on (thus only entry point into database is from applications server).
\\
\\
Finally, when considering keystores, it is important to have in mind that some implementations suffers from vulnerabilities. Several papers points out these vulnerabilities and contributes with proof-of-concept attacks to show the feasibility; for reference you may read \cite{sabt2016breaking}, \cite{focardi2018mind} and \cite{ryan2019hardware}.

\subsection{Solutions for high critical systems}
For highest security, given the assumption that networks are still needed and automated access to master key is allowed, following solution should be considered. The master key $k_{master}$ is stored in $\mathcal{E}$ (preferably a tamper-resistant, network separated HSM) with no key extraction ability, i.e. the key cannot be fully or partly extracted/exported. Furthermore, only operation that can be performed on the master key from an operational maintenance perspective is to compute the Key Check Value (KCV) in order to verify that the correct master key is present. Typical procedure to compute KCV is to send in the zero string for which $\mathcal{E}$ encrypts under the master key and outputs the cipher text truncated into 48 or 64 bytes.

Architectural setup is $n-1$ segmented subnetworks, all protected by standard means such as firewalls and gateways. Each subnet then contains one secret sharing party holding a share $k_i$ which is needed for access to $k_{master}$, i.e. creating an access key $k_n$ from the secure multi-party computation which unlocks $\mathcal{E}$ in order to access $k_{master}$. The requesting application $\mathcal{R}$ that needs an encryption or signature using $k_{master}$ triggers the multi-party computation by sending an authenticated and encrypted broadcast to all parties or via a facade API which in turn relays the request to all parties, which then securely sends each secret to HSM for final authentication. More precisely, $\mathcal{R}$ sends a request $(m,\omega)$ containing any data $m$ to be operated on and some authentication value $\omega$. Note that $\omega$ may be anything appropriate such as a signature on $m$ if public-key infrastructure is utilized, i.e. the usage of certificates, or it could be an access token if OpenID Connect infrastructure is used.
\\
\begin{figure}[H]
\centering
  \includegraphics[width=8cm]{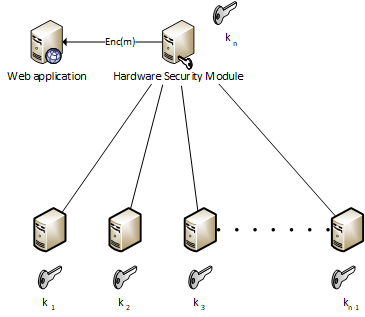}
  \caption{Secret sharing with multi-party computations using $n-1$ shares}
\end{figure}

Above simplified description requires some clarification on authentication details since it is hard to generalize too much and use cases may differ heavily. We consider two separate cases for the first flow when the request is sent:
\\
\begin{itemize}
\item{Public $\mathcal{R}$: a web application or public API which is exposed to an untrusted environment such as the Internet. In this case we could consider using TLS from the backend part of the API towards the SMPC part, but it does not handle the authenticity of the actual end-user. This would open up for an untrusted user to initiate the SMPC flow for any message $m$. If the use case does not allow this, a separate authentication layer is needed, e.g. OpenID Connect or some 2FA solution.}
\item{Private $\mathcal{R}$: a web application or API within an organization, behind a DMZ. In this case we assume any end-user involved is already authenticated earlier in the use case flow hence focus is on securing the actual message. Simplest solution is TLS.}
\end{itemize}
\;
\subsection{Security analysis}
\subsubsection*{Attack model}
To be more precise in the analysis of securing secret keys we need to setup a framework for how to consider the security of this particular use case; every attempt at securing the keys will suffer from one or several potential attacks. The attack vector is two dimensional, i.e. divided into power of adversary and type of attack surface. Let $\mathcal{R}$ be the requesting party, i.e. the intended user of using the master key; this could be a web application, system or even an end user. Next, let $\mathcal{E}$ be the secure storage entity, e.g. a vault, keystore or HSM. In a mathematical sense we could consider $\mathcal{R}$ sending some request message $m$ which triggers $\mathcal{E}$ to output a desired reply $c$, which can be a cipher text, a signature or a derived ephemeral key. Note that $m$ does not necessarily have to be sent directly to $\mathcal{E}$, but instead to one or several third parties which in turn trigger the computation in $\mathcal{E}$.
With increasing severity we order the attack model as follows:
\\
\subsubsection{Attacking secure channel}
Regardless of solution, may it be using secret sharing, multi-party computations or simply requesting keys or operations from $\mathcal{E}$, some sort of secure channel is needed between $\mathcal{R}$ and $\mathcal{E}$. This attack means that the adversary is eavesdropping on all network packets between $\mathcal{R}$ and $\mathcal{E}$. Naturally, if the secure channel is using TLS then the security is reduced to the security of TLS.
\\
\subsubsection{Attacking $\mathcal{R}$}
If the attacker is able to compromise $\mathcal{R}$ at some stage in order to send its own requests of choice to $\mathcal{E}$, possibly not even successful requests but at least able to trigger a communication flow, it is considered a successful attack. The base scenario is that the attacker constructs a request which $\mathcal{R}$ sends successfully to $\mathcal{E}$, regardless of content. This type of attack could reveal or leak information used for further exploits. If $\mathcal{R}$ is compromised on memory level then the adversary is able to extract any secret keys written into RAM. This implies that a master key should not be extracted from a vault or HSM but rather operating within that secure domain. What an adversary of this level can do is to replace or generate a request to use the master key of her choice; a significant difference compared to the attacking model described above.
\\
\subsubsection{Attacking $\mathcal{P}$ or $\mathcal{T}$}
Another attack possible is in a secret sharing and multi-party computational setting, where one or several third parties are involved; either some trusted third party $\mathcal{T}$ or a set of multi-party computational nodes $\mathcal{P} = \{P_1, ..., P_n\}$. If the attacker is able to read other parties' shares when $\mathcal{R}$ is collecting or handling them before calling $\mathcal{E}$, it implies that $\mathcal{R}$ should only send requests to $\mathcal{E}$ and not be part of any computations; instead $\mathcal{E}$ or other trusted parties, preferably in secure domains, initiate and perform necessary secure computations before accessing the master key. 
\\
\\
If $\mathcal{E}$ is compromised on memory level then there is no means of security left, thus the strongest attack possible.
\\
\\
One may consider investigate in certificateless cryptography as an alternative to TLS and certificates, but this would involve a trusted third party service and protocols based on pairings which may be slower in performance. However, in a recent proof of concept of which Java was used as underlying platform, some evidence of practical usage of certificateless cryptography for securing master keys has been shown \cite{fokin2018secure}.

\section{Conclusion}
It seems like much effort and high complexity of securing a master key in this way; the temptation to just leave it as it is and store the key embedded or in a database is understandable. Also, much of the presented cryptography in this paper has not yet reached a level of confidence nor consensus in industry and that is why we highlight them now. As conjured in the paper's title we assume this is an unsolvable problem but the point is to align towards the security need by making it as hard as possible for an adversary to compromise a master key. By using non-traditional approaches such as multi-party computations as proposed, our analysis is that it would not be too complex to utilize such technology and the security benefits would balance out the effort and time of implementation.

\section{Future work}
Considering the secure channel to be used between parties $P_i$ and $\mathcal{E}$, not only do we have secure segmentation between subnets but we would also like to add a layer of security in case messages are eavesdropped or tampered with. For this we propose a few different approaches that can be explored:
\\
\begin{itemize}
\item{Public-key infrastructure, i.e. use of traditional private-public keypairs for which each $P_i$ uses to connect to HSM server via TLS. Drawback is high maintenance of certificates and some risk associated to TLS vulnerabilities \cite{sirohi2016comprehensive}}
\item{Zero-Knowledge Proof (ZKP) based authentication, i.e. setting up a protocol where the HSM stores a secret \emph{claim} (some data $c_i$) which is unique for each party $P_i$. Then each $P_i$ initiates a ZKP protocol and proves that they know $c_i$ without revealing any data; by doing so the HSM can trust that $P_i$ is eligible to contribute with a share $k_i$.}
\item{Certificateless cryptography, as mentioned above having a trusted third party $\mathcal{T}$ within the secure domain of the network which handles the keys needed for encryption and authentication. Many solutions builds on pairings which may not be efficient enough depending on requirements.}
\end{itemize}
\;\;
Regardless of communication protocols chosen we also see that more research should be put into the domain of securing secret keys, especially in untrusted and dishonest environments. Not only the encryption part but rather schemes or methods for securely retrieving sensitive data without leakage; perhaps one should investigate more in secret data sets, secret sharing, ZKP schemes and similar cryptography. We could also expect more use cases where such secret key retrieval is needed but in environments with distributed computing (cloud) and connected devices (IoT).






%


\bibliographystyle{plain}
\bibliography{references}

\end{document}